\documentclass[journal]{IEEEtran}
\usepackage[noadjust]{cite}
\usepackage[english]{babel}
\usepackage{graphicx}
\usepackage{tikz}
\usepackage{circuitikz}
\usepackage{xcolor}
\usepackage{picinpar}
\usepackage{amsmath}
\usepackage{amsthm}
\usepackage{amssymb}
\usepackage{mathrsfs}
\usepackage{booktabs}
\usepackage{tabularx}
\usepackage{pifont}   
\usepackage{graphicx} 
\usepackage{caption}
 \usepackage{pifont}
 \usepackage{enumitem}
\usepackage[T1]{fontenc}
\usepackage{url}
\usepackage{inputenc}
\usepackage{makecell}
\usepackage{booktabs}
\usepackage{colortbl}
\usepackage{soul}
\usepackage{multirow}
\usepackage{eso-pic}
\usepackage{tcolorbox}
\usepackage{pifont}
\usepackage{color}
\usepackage{alltt}
\usepackage[hidelinks]{hyperref}
\usepackage{tabularx}
\usepackage{siunitx}
\usepackage{epstopdf}
\usepackage{subcaption}
\usepackage{lipsum}
\usepackage{array}
\usepackage{stfloats}
\usepackage{pbox}
\usepackage{float}
\usepackage{ragged2e}
\usepackage{amssymb}
\usepackage[english]{babel}
\usepackage{caption}
\usepackage{lettrine}
\usepackage{lipsum} 
\usepackage{etoolbox}
\makeatletter

\newcommand{\introductioncap}[1]{%
  \lettrine[lines=2, lhang=0.3, nindent=0em, findent=0em]{\uppercase{#1}}{}%
}
\makeatletter
\renewcommand{\fnum@figure}{Fig. \thefigure}
\makeatother

\begin{document}
 \title{Multi-Snapshot Deep Denoising for Channel Estimation in OTFS Modulated Systems}
 \author{
Surbhi Gehlot, \textit{Graduate Student Member, IEEE},
Siddhi Shinde,
Suraj Srivastava, \textit{Member, IEEE},
and Sandeep Kumar Yadav

\thanks{
Surbhi Gehlot, Siddhi Shinde, and Sandeep Kumar Yadav are with the Department of Electrical Engineering, Indian Institute of Technology Jodhpur, Jodhpur 342030, India
(e-mail: gehlot.5@iitj.ac.in; m24cps008@iitj.ac.in; sy@iitj.ac.in).
}

\thanks{
Suraj Srivastava is with the Department of Electronics Engineering, Indian Institute of Technology Jodhpur, Jodhpur 342030, India
(e-mail: surajsri@iitj.ac.in).
}

\thanks{
The work of S. Srivastava was supported in part by IIT Jodhpur's Research Grant I/RIG/SUS/20240043; in part by Anusandhan National Research Foundation under Grant PM-ECRG/2024/478/ENS and Grant ANRF/ARG/2025/005895/ENS; in part by Telecom Technology Development Fund (TTDF) under Grant TTDF/6G/368; in part by ICON-Project through the India Department of Science and Technology and in part by UKRI–EPSRC under India–U.K. Joint opportunity in
Telecommunications Research
}
}
 
\maketitle
\AddToShipoutPictureFG*{%
  \AtPageLowerLeft{%
    \ifnum\value{page}=1
    \hspace*{0.1\textwidth}%
    \raisebox{0.2cm}{%
          \begin{tcolorbox}[colback=gray!10, colframe=black, width=0.95\textwidth, boxrule=0.5pt]
      \footnotesize
      This paper has been published in the IEEE Communications Letters.
      The article is available at IEEE Xplore: \href{https://ieeexplore.ieee.org/document/11523444} {\underline{10.1109/LCOMM.2026.3694255}}.
      \end{tcolorbox}
    }%
    \fi
  }%
}	
\begin{abstract}
A deep denoising based channel estimation framework is proposed for orthogonal time frequency space (OTFS) modulated systems, wherein channel state information (CSI) recovery is formulated as an image restoration problem. A salient attribute of the approach is the exploitation of structural invariance in the delay Doppler (DD) domain channel over a geometric coherence time, allowing multiple OTFS frames captured during this period to serve as noisy snapshots of the approximately identical channel. These snapshots jointly enhance the effectiveness of the proposed lightweight denoiser based on nonlinear activation free network (NAFNet). The method exhibits low computational complexity, operates reliably even at low pilot signal-to-noise ratio (PSNR), and can accommodate both fractional delay and fractional Doppler effects. Simulation results demonstrate significant performance gains over the existing methods.

\end{abstract}


\section{Introduction}
\introductioncap{O}wing to its superior performance in high-mobility environments, orthogonal time frequency space (OTFS) has emerged as a strong candidate for next-generation wireless systems \cite{hadani2017otfs}. To harness the full benefit of OTFS, accurate estimation of the underlying channel state information (CSI) becomes imperative. 
Foundational methods employed a high-power pilot embedded within the data frame, surrounded by guard symbols to suppress interference \cite{raviteja2018practical, mehrotra2022channel}. Although more efficient, it relies on thresholding that remains highly sensitive to pilot signal-to-noise ratio (PSNR). Subsequent methods leveraged the inherent sparsity of the delay Doppler (DD) domain channel and employed compressed sensing (CS) techniques such as sparse Bayesian learning (SBL) \cite{srivastava2021bayesian, zhao2020sparse} and orthogonal matching pursuit (OMP) \cite{srivastava2022delay} for accurate CSI estimation. While these approaches achieved high performance under ideal sparsity conditions, their effectiveness deteriorates in the presence of fractional delay-fractional Doppler (FDFD). Moreover, to cater for the effects of FDFD, complexity increases significantly with finer DD resolution, limiting their practicality in real-world OTFS systems. Recently, owing to their ability to learn complex nonlinear mappings and enable fast inference through offline training, deep learning (DL) models have emerged as efficient tool for channel estimation \cite{soltani2019deep}. Several recent studies have explored DL for OTFS channel estimation by formulating the DD response as a denoising problem. Works in \cite{he2023denoising, li2022residual} employed OMP based coarse estimation followed by convolutional neural networks (CNN) based denoisers, while \cite{zhang2024sparse} integrated sparsity priors into a learned denoising model. In \cite{jing2024learned}, to recover the DD domain channel, a sparse adaptive estimator was coupled with fast deep video denoising network (FastDVDNet), a video denoising network. The use of video denoisers increases complexity and training overhead. Also, the above mentioned methods assume integer DD grids and rely on strong sparsity, which deteriorates under off-grid conditions. To alleviate this, \cite{guo2024otfs} proposed a super-resolution convolutional neural network (SRCNN) based framework to reconstruct CSI from low-resolution DD representations under fractional Doppler. However, fractional delays, equally critical, further spread energy and induce inter-path interference (IPI), challenging existing designs. Although some conventional works \cite{khan2023low,lei2025low} have addressed both FDFD effects via path-wise estimation, they depend on handcrafted thresholds and successive interference cancellation (SIC), which is prone to residual interference buildup. To address these limitations, our proposed learning-driven OTFS channel estimation framework contributes:
    \begin{enumerate}[nosep]
    \item The CSI estimation problem is formulated as an image restoration problem, thereby enabling the use of image denoising networks. Nonlinear activation-free network (NAFNet), a lightweight architecture, is adopted for its architectural simplicity and effective denoising capability.
    \item The proposed method exploits the structural invariance of the DD domain channel within a geometric channel coherence time by aggregating multiple noisy DD snapshots of the approximately identical channel in this interval. These observations are processed through the denoiser, followed by frame-wise averaging of the denoised outputs to further improve robustness, especially under low PSNR.
    \item The proposed framework eliminates manually tuned thresholds, thereby reducing PSNR sensitivity. It also removes coarse-to-fine denoising pipelines by learning the DD domain mapping directly from raw observations, thereby avoiding error propagation from coarse estimates and enabling a simpler, more robust CSI estimator.
    \item To ensure spectral efficiency and mitigate pilot–data interference, a dedicated embedded pilot structure is incorporated into the transmitted OTFS frame. Furthermore, for effective data detection, the estimated DD domain channel is leveraged by a minimum mean square error (MMSE) receiver.
    \end{enumerate}
 \textbf{Notation:} $\mathrm{vec}(\mathbf{A})$ and $\mathrm{vec}^{-1}(\cdot)$ denote vectorization and its inverse. $\|\cdot\|$, $\otimes$, $(\cdot)^H$, $\mathbb{E}[\cdot]$, and $\odot$ represent the Frobenius norm, Kronecker product, Hermitian transpose, expectation, and element-wise multiplication.

\section{OTFS System Description}
\label{OTFS System Description}
An OTFS system with bandwidth $B=M\Delta f$ and frame duration $T_f=NT$ is considered, where $N$ and $M$ are the information symbols along the Doppler and delay bins respectively. The DD domain is discretized into an $M\times N$ grid with delay and Doppler resolutions $\Delta\tau=1/B$ and $\Delta\nu=1/T_f$, respectively. Let $\mathbf{X}_{\mathrm{DD}}\in\mathbb{C}^{M\times N}$ denote the DD domain symbol matrix, whose $(l,k)$-th entry $\mathbf{X}_{\mathrm{DD}}(l,k)$ corresponds to the symbol transmitted at  delay index $l$ and Doppler index $k$. OTFS modulation maps $\mathbf{X}_{\mathrm{DD}}$ to the time--frequency (TF) domain using the inverse symplectic finite Fourier transform (ISFFT), i.e., $\mathbf{X}_{\mathrm{TF}}=\mathbf{F}_M\mathbf{X}_{\mathrm{DD}}\mathbf{F}_N^{{H}}$, where $\mathbf{F}_M\in\mathbb{C}^{M\times M}$ and $\mathbf{F}_N\in\mathbb{C}^{N\times N}$ are discrete Fourier transform (DFT) matrices. The TF-domain symbols are converted to discrete-time samples via the Heisenberg transform. With transmit pulse shaping, the resulting time domain (TD) signal matrix is $\mathbf{S}=\mathbf{G}_{\mathrm{tx}}\mathbf{X}_{\mathrm{DD}}\mathbf{F}_N^{{H}}$, where $\mathbf{G}_{\mathrm{tx}}=\mathrm{diag}\{g_{\mathrm{tx}}(mT/M)\}_{m=0}^{M-1}$ and $g_{\mathrm{tx}}(t)$ is a pulse shaping of duration $T$. Vectorizing $\mathbf{S}$ yields $\mathbf{s}=(\mathbf{F}_N^{{H}}\otimes\mathbf{G}_{\mathrm{tx}})\mathbf{x}_{\mathrm{DD}}$, with $\mathbf{x}_{\mathrm{DD}}=\mathrm{vec}(\mathbf{X}_{\mathrm{DD}})$. A cyclic prefix (CP) of length $N_{\mathrm{cp}}$ is appended to eliminate inter-symbol interference caused by multipath delay spread \cite{raviteja2018practical}.

The wireless channel consists of $P$ propagation paths, where the complex gain, delay, and Doppler shift of the $p$-th path are denoted by $h_p$, $\tau_p$, and $\nu_p$, respectively. The corresponding DD domain channel representation is given by
$
h(\tau,\nu)=\sum_{p=1}^{P} h_p\,\delta(\tau-\tau_p)\delta(\nu-\nu_p).
$
In practical OTFS systems, the delay and Doppler shifts are not restricted to integer multiples of the DD-grid resolutions and are expressed as $\tau_p=(l_p+\iota_p)/(M\Delta f)$ and $\nu_p=(k_p+\kappa_p)/(NT)$, where $l_p,k_p\in\mathbb{Z}$ and $\iota_p,\kappa_p\in(-\tfrac{1}{2},\tfrac{1}{2})$ denote the fractional delay and Doppler offsets \cite{lei2025low}.
After CP removal, the received signal can be expressed as $\mathbf{r}=\mathbf{H}\mathbf{s}+\mathbf{w} $ where $\mathbf{r}\in\mathbb{C}^{MN\times 1}$ and
 $\mathbf{w}\in\mathbb{C}^{MN\times 1}$
denote the received signal and the
noise process, respectively. Also, the effective
TD channel matrix is given by
$\mathbf{H}=\sum_{p=1}^{P} h_p\,\boldsymbol{\Pi}_{\tau_p}\boldsymbol{\Delta}_{\nu_p}$.
Here, $\boldsymbol{\Pi}_{\tau_p}$ denotes a circular delay operator corresponding to
the propagation delay $\tau_p$, while $\boldsymbol{\Delta}_{\nu_p}$ denotes a
Doppler modulation operator associated with the Doppler shift $\nu_p$. 

At the receiver, the TD signal $\mathbf{r}$ is converted to the DD
domain via OTFS demodulation. Let $\mathbf{R}=\mathrm{vec}^{-1}(\mathbf{r})\in
\mathbb{C}^{M\times N}$ be the received sample matrix. The TF-demodulated
signal $\mathbf{Y}_\mathrm{TF}\in\mathbb{C}^{M\times N}$ is obtained by applying the
discrete Wigner transform as $\mathbf{Y}_\mathrm{TF}=\mathbf{F}_M\mathbf{G}_{\mathrm{rx}}\mathbf{R}$,
where $\mathbf{G}_\mathrm{rx}=\mathrm{diag}\{g_\mathrm{rx}^*(mT/M)\}_{m=0}^{M-1}$ represents the
received pulse-shaping filter of duration $T$. The DD domain signal
$\mathbf{Y}_\mathrm{DD}\in\mathbb{C}^{M\times N}$ is then obtained via the symplectic finite
Fourier transform (SFFT) as $\mathbf{Y}_\mathrm{DD}=\mathbf{F}_M^{{H}}\mathbf{Y}_\mathrm{TF}
\mathbf{F}_N$, with vectorized form $\mathbf{y}_\mathrm{DD}=\mathrm{vec}(\mathbf{Y}_\mathrm{DD})$.
Substituting the TD input--output (IO) relation yields the DD domain model $
\mathbf{y}_\mathrm{DD}=\mathbf{H}_\mathrm{DD}\mathbf{x}_\mathrm{DD}+\mathbf{v}_\mathrm{DD},
$
where $\mathbf{v}_{\mathrm{DD}}\sim\mathcal{CN}(\mathbf{0},\sigma^2\mathbf{I}_{MN})$ denotes 
circularly symmetric complex Gaussian noise process with zero mean and variance $\sigma^2$, and
$\mathbf{H}_{\mathrm{DD}}=(\mathbf{F}_N\otimes\mathbf{G}_{\mathrm{rx}})\mathbf{H}
(\mathbf{F}_N^{H}\otimes\mathbf{G}_{\mathrm{tx}})$.
 $\mathbf{H}_{\mathrm{DD}}\!\big[(l',k'),(l,k)\big]$ can also be expressed as 
 $\sum_{p=1}^{P} h_p\,\beta(l',k')\,
\alpha_{l',k'}(\tau_p,\nu_p)$. Equivalently, the DD domain IO relation is expressed as

\begin{align}
\mathbf{Y}_{\mathrm{DD}}(l',k')
&=
\sum_{l=0}^{M-1}\sum_{k=0}^{N-1}
\mathbf{X}_{\mathrm{DD}}(l,k)
\sum_{p=1}^{P} h_p\,
\beta(l',k')\,
\alpha_{l',k'}(\tau_p,\nu_p)
\nonumber\\
&\quad + \mathbf{V}_{\mathrm{DD}}(l',k') ,
\label{eq:dd_elementwise}
\end{align}
The term $\alpha_{l',k'}(\tau_p,\nu_p)$ captures the deterministic
two-dimensional spreading induced by the fractional delay $\tau_p$ and
Doppler shift $\nu_p$ of the $p$-th path \cite{khan2023low},
and is given by

\begin{equation}
\begin{aligned}
\alpha_{l',k'}(\tau_p,\nu_p)\footnotemark[1]
&=
\underbrace{\frac{1}{N}\sum_{n=0}^{N-1}
e^{-j2\pi n\left(\frac{k'-k}{N}-\nu_p T\right)}}_{\text{due to Doppler-domain spreading}} \\
&\quad \times
\underbrace{\frac{1}{M}\sum_{m=0}^{M-1}
e^{j2\pi m\left(\frac{l'-l}{M}-\tau_p \Delta f\right)}}_{\text{due to delay-domain spreading}} .
\end{aligned}
\label{eq:alpha_kernel}
\end{equation}

\footnotetext[1]{%
For integer delay--Doppler shifts, 
$\alpha_{l',k'}(\tau_p,\nu_p)$ reduces to a single dominant coefficient at
$\big([l+l_p]_M,\,[k+k_p]_N\big)$.
For $\nu_p = 0$ (or $\tau_p = 0$), spreading occurs only along the delay
(or Doppler) dimension.}

This kernel describes how the energy of a single propagation path spreads
over neighbouring DD domain indices due to fractional Doppler (along $k$)
and fractional delay (along $l$), resulting in a generally dense
effective DD domain response. The term $\beta(l',k')$
represents a per path deterministic phase term \cite{khan2023low}. Further, in the simplified DD domain IO relation in \eqref{eq:dd_elementwise}, for a pilot placed at the DD index $(0,0)$, the per-path deterministic phase  becomes a constant \cite{mehrotra2022channel}, so can be absorbed into the complex path gain $h_p$.

\section{Proposed Framework}
\label{CSI Estimation Model for OTFS Systems}
An embedded DD domain pilot forms noisy multi-frame observations under FDFD spreading. These snapshots are jointly processed by an image restoration network to estimate the effective DD domain channel, which is subsequently used for MMSE data detection.
\subsection{Pilot Observation and Effective Channel Formation}
\label{pilot}
\begin{figure}[htb]
    \centering
\includegraphics[height=5cm]{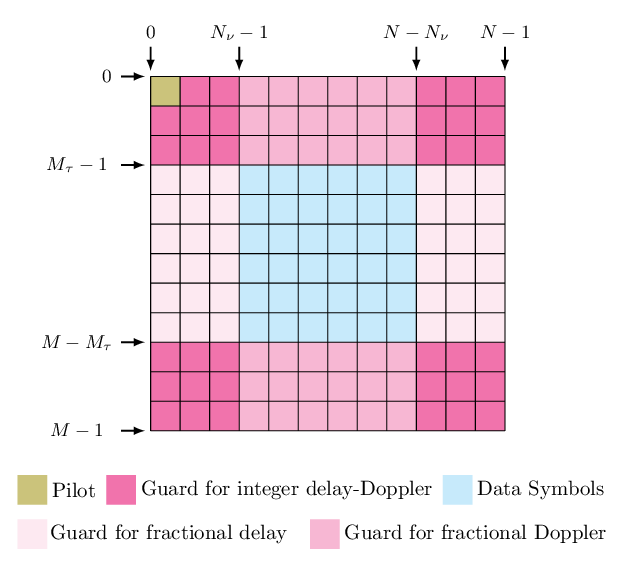}
    \caption{Proposed frame architecture.}
    \label{frame}
\end{figure}

An embedded impulse pilot is placed in the DD domain to form observations of the effective channel response. The transmitted DD domain symbol matrix is defined as

\begin{equation}
\mathbf{X}_{\mathrm{DD}}(\ell,k)=
\begin{cases}
\sigma_p, & (\ell,k)=(0,0), \\[2pt]
\text{guard}, & (\ell,k)\in\mathcal{G}_{\mathrm{int}} \cup \mathcal{G}_{\tau} \cup \mathcal{G}_{\nu}, \\[2pt]
\text{data symbol}, & \text{otherwise},
\end{cases}
\label{eq:xdd}
\end{equation}
 where $\sigma_p$ is the pilot amplitude. In CP-aided OTFS, the circular convolution causes the DD domain response of the pilot at $(0,0)$ to spread across the grid, with significant components appearing at boundaries of the DD frame. Therefore, to suppress pilot–data interference due to both integer and fractional DD components, the DD grid is partitioned into the following guard regions: $\mathcal{G}_{\mathrm{int}} = \{(\ell, k)\,|\, \ell \in [0{:}M_\tau{-}1] \cup [M{-}M_\tau{:}M{-}1],\; k \in [0{:}N_\nu{-}1] \cup [N{-}N_\nu{:}N{-}1]\}$ covers the four integer-interference corners, $\mathcal{G}_\nu = \{(\ell, k)\,|\, \ell \in [0{:}M_\tau{-}1] \cup [M{-}M_\tau{:}M{-}1],\; k \in [N_\nu{-}1{:}N{-}N_\nu]\}$ for fractional Doppler spread, and $\mathcal{G}_\tau= \{(\ell, k)\,|\, \ell\in[M_\tau{-}1{:}M{-}M_\tau],\; k \in [0{:}N_\nu{-}1] \cup [N{-}N_\nu{:}N{-}1]\}$ for fractional delay spread, as depicted in Fig. \ref{frame}
 
Further, the physical channel parameters $\{h_p, \tau_p, \nu_p\}_{p=1}^{P}$ remain invariant over the multiple OTFS frames, as the DD impulse response is governed by the underlying propagation geometry and evolves significantly more slowly than the time-varying channel representation \cite{hadani2017otfs}. Consequently, the effective DD domain channel response can be treated as invariant across multiple OTFS frames within the geometric coherence interval. Let $F$ snapshots be observed within this interval. Then, the $f$-th DD domain pilot observations across frames satisfy
\begin{equation}
\mathbf{Y}_{\mathrm{DD}}^{(f)}(l',k')
=
\sigma_p \sum_{p=1}^{P} h_p\, \alpha_{l',k'}(\tau_p,\nu_p)
+
\mathbf{V}_{\mathrm{DD}}^{(f)}(l',k'),
\end{equation}
where $\mathbf{V}_{\mathrm{DD}}^{(f)}(l',k')$ is independent and identically distributed
across frames.
After normalization by \(\sigma_p\), each frame i.e. $\mathbf{Y}_{\mathrm{DD}}^{(f)}(l',k')/\sigma_p,$ provides a noisy snapshot of the same underlying effective
DD domain channel response
$\mathbf{H}_{\mathrm{eff}}(l',k')
\triangleq
\sum_{p=1}^{P} h_p\, \alpha_{l',k'}(\tau_p,\nu_p)$. Since each frame provides an independent noisy observation of the same effective DD domain channel realization, the frame-wise estimates are combined through simple averaging,
$\widehat{\mathbf{H}} = \frac{1}{F}\sum_{f=1}^{F}\widehat{\mathbf{H}}_f$,
which corresponds to the classical sample mean estimator for independent Gaussian measurements and achieves the minimum variance among unbiased estimators.
Consequently, effective DD domain channel estimation reduces to recovering
\(\mathbf{H}_{\mathrm{eff}}(l',k')\) from multiple noisy observations. These multi-frame snapshots are jointly exploited
by the proposed denoising-based channel estimation framework described next.

\subsection{Denoising-Based Channel Estimation}
\label{csi estimation}
Let $\mathbf{Z}_f \in \mathbb{C}^{M_\tau \times N_\nu}$ denote the DD domain pilot observation corresponding to the $f$-th OTFS frame. It is mapped to a real-valued two-channel feature map of size $M_\tau \times N_\nu \times 2$, given by $\big[\Re\{\mathbf{Z}_f\},\,\Im\{\mathbf{Z}_f\}\big]$. In practical OTFS channels, the effective DD support satisfies 
$M_\tau \ll M$ and $N_\nu \ll N$ \cite{ srivastava2021bayesian}, resulting in a compact feature map of size $M_\tau \times N_\nu$. 
For such small representations, employing very deep restoration networks designed for high-resolution inputs would introduce unnecessary architectural complexity. NAFNet is particularly suitable in this setting as it preserves the DD domain resolution by avoiding spatial size reducing operations such as pooling, while maintaining a lightweight architecture.\\
Each frame corresponds to the same underlying physical channel realisation and is processed independently by the denoiser. The denoising operation is implemented using NAFNet \cite{chen2022simple}, denoted by $\mathcal{F}_\theta(\cdot)$ with learnable parameters $\theta$. For each OTFS frame, the denoised DD domain channel estimate is given by

\begin{figure}[htb]
    \centering
\includegraphics[width=\linewidth,height=8.15cm]{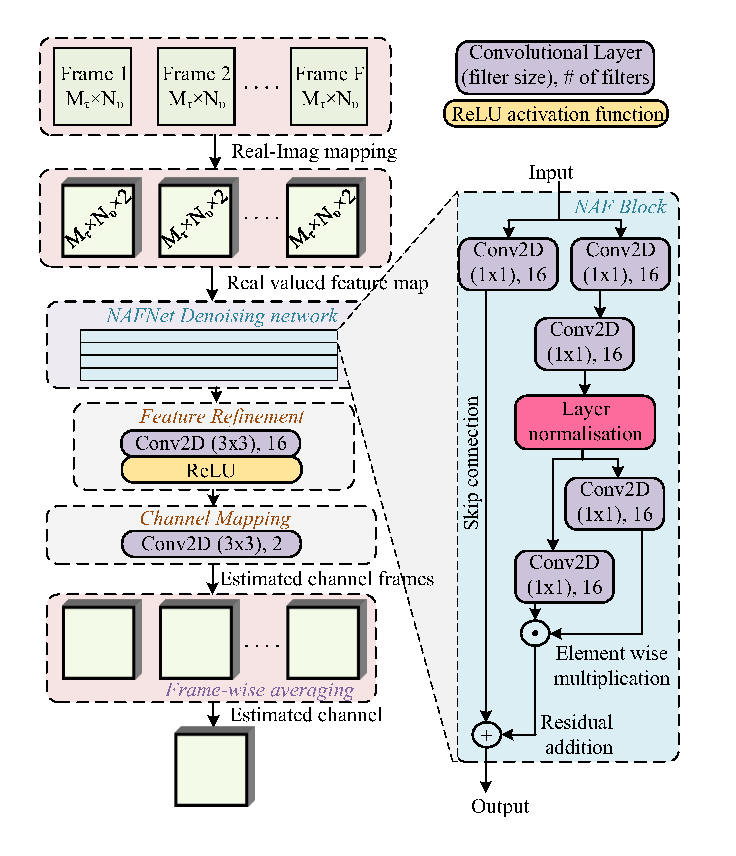}
    \caption{Proposed NAFNet based image restoration framework.}
    \label{nafnet}
\end{figure}
\begin{equation}
\widehat{\mathbf{H}}_f = \mathcal{F}_\theta\!\left( \mathbf{Z}_f \right), \quad f = 1,2,\ldots,F.
\end{equation}

As illustrated in Fig. \ref{nafnet}, the NAFNet architecture consists of a cascade of $B$ identical processing blocks. Each block implements a residual denoising operation of the form $\mathcal{B}(\mathbf{X}) = \mathbf{A}(\mathbf{X}) + \mathcal{G}(\mathbf{X})$ where $\mathbf{A}(\cdot)$ denotes a projection skip connection that aligns feature dimensions using a $1\times1$ pointwise convolution. The residual branch $\mathcal{G}(\cdot)$ employs multiplicative gating, expressed as \begin{equation}
\mathcal{G}(\mathbf{X}) = \mathbf{U}(\mathbf{X}) \odot \mathbf{V}(\mathbf{X}).
\label{eq:naf_gating}
\end{equation} 
 
The transformations $\mathbf{U}(\cdot)$ and $\mathbf{V}(\cdot)$ share identical convolutional pipelines comprising a $1\times1$ pointwise convolution, a $3\times3$ spatial convolution, and layer normalization. The $1\times1$ operation transforms features independently at each DD bin, while the $3\times3$ operation captures neighborhood structure from FDFD induced energy spread. Their combination enables successive enhancement of DD domain features across the network depth, while layer normalization mitigates scale imbalance across feature channels prior to multiplicative gating. Also, no explicit nonlinear activation functions are employed within the block; instead, nonlinearity arises implicitly via multiplicative gating in the residual branch \eqref{eq:naf_gating} \cite{chen2022simple}. After the cascade of processing blocks, a final convolutional refinement stage consolidates the learned DD domain features and is followed by a linear projection that maps the representation back to the DD domain channel. This produces the set of frame-wise channel estimates $\{\widehat{\mathbf{H}}_f\}_{f=1}^{F}$. As the effective DD domain channel is invariant across frames, multi-frame averaging is applied to further suppress residual noise. The network parameters are learned by minimizing the mean-squared error (MSE) between the estimated and true DD domain channels, i.e., $\mathcal{L}(\theta)=\mathbb{E}\!\left[
\left\|
\widehat{\mathbf{H}}_f - \mathbf{H}_{\mathrm{true}}
\right\|^{2}
\right].$ The resulting DD domain channel estimate is then used for linear data detection.

\subsection{Linear Data Detection Using Estimated DD domain CSI}
 \label{data detetction}
 
Following the embedded pilot–guard frame structure defined in \eqref{eq:xdd}, data symbols occupy the DD domain locations not occupied by the pilot or guard symbols. Let $\mathbf{x}_\mathrm{DD}$ denote the transmitted DD domain data symbols. Let, the estimated channel constructed from the multi-frame averaged denoised output be $\widehat{\mathbf{H}}_{\mathrm{DD}}\in\mathbb{C}^{MN\times MN}$. 
The resulting DD domain IO relationship corresponding to the data symbols is given by $
\mathbf{y}_{\mathrm{DD}}
=
\widehat{\mathbf{H}}_{\mathrm{DD}},
\mathbf{x}_{\mathrm{DD}}
+
\mathbf{v}_{\mathrm{DD}},
$
where $\mathbf{v}_{\mathrm{DD}}\sim\mathcal{CN}(\mathbf{0},\mathbf{I})$ denotes additive circularly symmetric complex Gaussian noise. The DD domain channel estimate used for data detection is obtained from the multi-frame averaged denoised channel estimate described in Section \ref{csi estimation}. Assuming data symbols of average power $\sigma_d^2$, linear data detection is performed using an MMSE detector constructed from the estimated DD domain CSI as $\big(
\widehat{\mathbf{H}}_\mathrm{DD}^{H}
\widehat{\mathbf{H}}_\mathrm{DD}
+
\tfrac{1}{\sigma_d^{2}}\mathbf{I}
\big)^{-1}
\widehat{\mathbf{H}}_{\mathrm{DD}}^{H}
\mathbf{y}_\mathrm{DD} .$

\begin{table}[!t]
\footnotesize
\renewcommand{\arraystretch}{1} 
\caption{Simulation parameters}
\label{tab:simulation}
\centering
\label{sub6parameters}
\begin{tabular}{|l|l|}
\hline
\multicolumn{1}{|c|}{\textbf{Parameter (symbol)}} & \multicolumn{1}{c|}{\textbf{Value}} \\ \hline
Carrier Frequency in GHz ($f_c$) & $4$ \\ \hline
Subcarrier spacing in kHz ($\Delta f$) & $15$ \\ \hline
No. of symbols along delay-axis ($M$) & $32$ \\ \hline
No. of symbols along Doppler-axis ($N$) & $32$ \\ \hline
Maximum user speed in km/h ($v_{max}$)  &   507.6\\
\hline
Max. spread across delay-axis ($M_\tau$) & $8$ \\ \hline
Max. spread across Doppler-axis ($N_\nu$) & $8$ \\ \hline

Modulation scheme & BPSK\\ \hline
No. of dominant reflectors ($P$) & $5$ \\ \hline

No. of training snapshots (per PSNR) & $6000$ \\ \hline

No. of frames ($F$) & $5$  \\ \hline

\end{tabular}
\end{table}

\begin{figure*}[!htb]
    \centering 
        \begin{subfigure}{0.22\textwidth}
 
  \includegraphics[width=4.25cm,height=4.19cm]{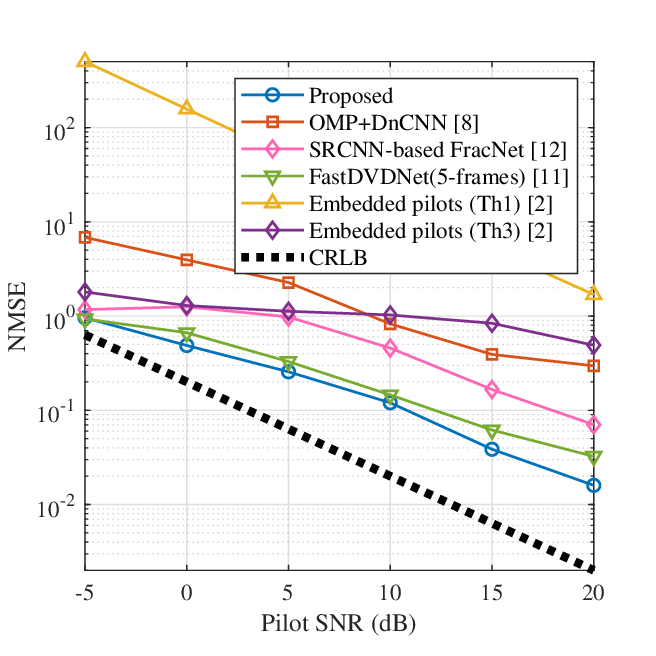}
	
    \caption{NMSE vs Pilot SNR (PSNR).}
	\label{nmse comparison}
\end{subfigure}\hfil 
\begin{subfigure}{0.22\textwidth}
  \includegraphics[width=4.25cm,height=4.19cm]{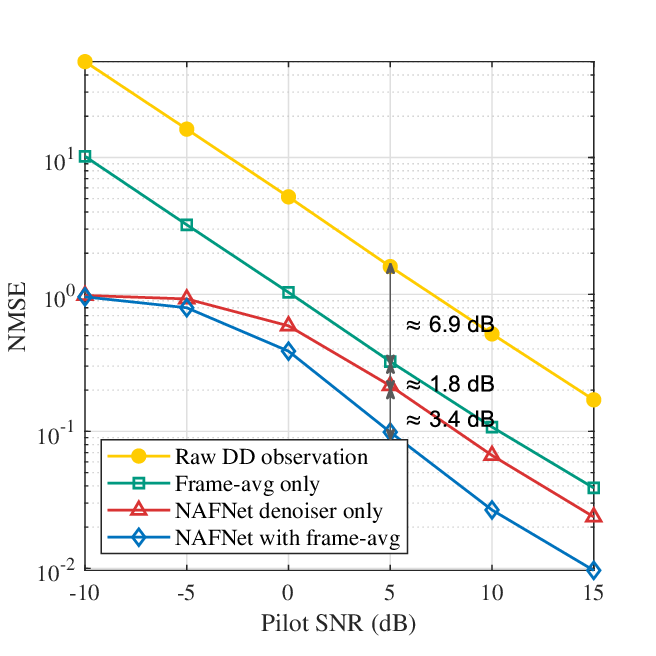}
	\caption{NAFNet and frame-averaging.}
        \label{frames}
\end{subfigure}\hfil
\begin{subfigure}{0.22\textwidth}

 \includegraphics[width=4.25cm,height=4.19cm]{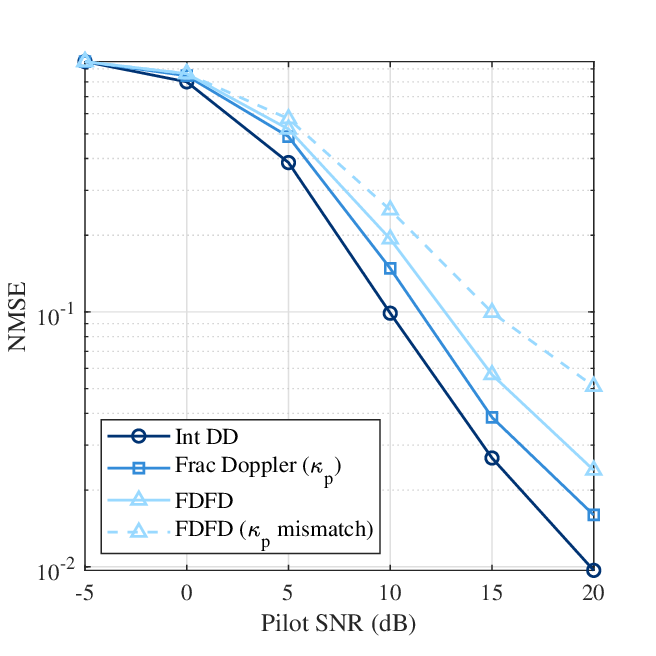}
	\caption{Fractional DD robustness.}
	\label{frac effect}
\end{subfigure}\hfil
\begin{subfigure}{0.22\textwidth}
  \includegraphics[width=4.21cm,height=4.14cm]{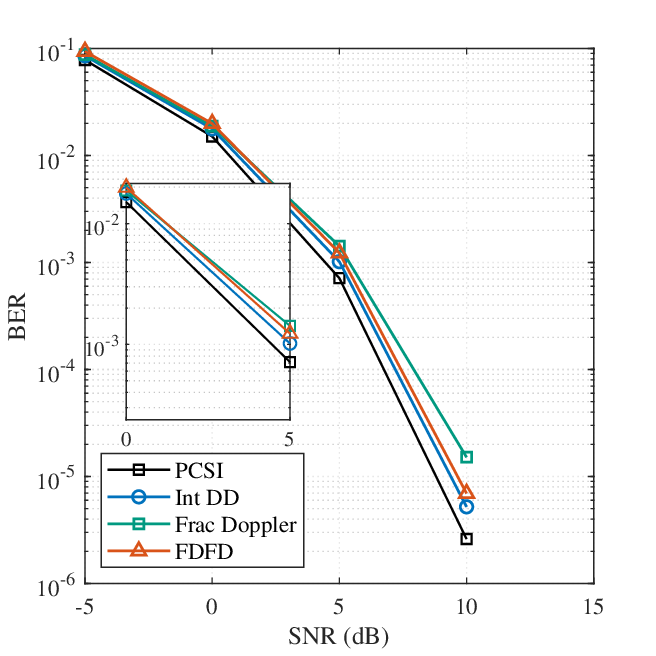}
	\caption{BER performance vs data SNR.}
        \label{ber}
\end{subfigure}

 \caption{Performance analysis of the proposed framework.} 
        \label{Fig_Results}
\end{figure*}

\section{Results and Discussion}
\label{Results and Discussion}
The simulations follow the parameter settings summarised in Table \ref{tab:simulation}. A high-mobility DD channel profile is adopted following the structured multipath model in \cite{srivastava2021bayesian}. Based on the system parameters, all simulated frames span a duration of $F T_f \approx 10.6$ ms, corresponding to a maximum displacement of $\approx 1.5$ m at $v_{\max}$. This displacement is negligible relative to typical propagation distances, thereby ensuring that the DD parameters remain invariant across $F$ OTFS frames. To reflect practical propagation effects, each DD tap location is extended to include sub-grid delay and Doppler shifts drawn from \( \mathcal{U}(-0.5, 0.5) \). The dataset is partitioned into training, validation, and testing subsets in a $6{:}2{:}2$ ratio. The NAFNet architecture employs $4$ residual blocks with a batch size of $64$ and is trained using the Adam optimizer. Key hyperparameters are selected using automated hyperparameter optimization\textsuperscript{2}.\\
The performance of the proposed framework is evaluated using the normalized mean square error (NMSE) and the symbol error rate (SER). For a given channel realisation, the NMSE is defined as $\mathrm{NMSE} = \|\widehat{\mathbf{H}}_{\mathrm{DD}} - \mathbf{H}_{\mathrm{DD}}\|^2 / \|\mathbf{H}_{\mathrm{DD}}\|^2$. Fig. \ref{nmse comparison} presents the NMSE performance of the proposed method versus PSNR. For comparison, an embedded pilot based method \cite{raviteja2018practical}, an OMP based denoising framework \cite{he2023denoising}, SRCNN-based FracNet model \cite{guo2024otfs} and FastDVDNet based video denoiser with 5 frames \cite{jing2024learned} are considered. The embedded pilot approach achieves its best performance at an optimal threshold of $3$, while its performance at other threshold settings degrades significantly. The OMP based method is vulnerable to support detection errors. Further, both FastDVDNet and FracNet incur higher inference time, while the latter additionally relies on a threshold dependent coarse estimate. In contrast, the proposed framework directly learns a DD domain mapping from raw DD domain observations, providing clear gains in the low PSNR regimes. Also, as the PSNR increases, the estimation problem becomes increasingly well-conditioned, leading to the gradual convergence of the NMSE curves across all schemes.

\footnotetext[2]{Hyperparameter optimisation is performed using the Optuna framework. Early stopping is applied within each trial to improve generalisation and reduce computational overhead.}

Next, Fig. \ref{frames} presents an ablation study with four cases: raw DD observations, frame averaging of the raw DD frames, NAFNet denoising using a single frame, and frame averaging applied to the NAFNet denoised estimates. Five frames are used for averaging in both the baseline frame-averaging scheme and when averaging is applied to the denoised outputs. While averaging the raw DD frames reduces noise across frames, the NAFNet denoiser significantly improves the channel estimate even from a single frame. Morever, note that when the denoised estimates are further averaged across frames, an additional NMSE reduction is observed, indicating that the two operations provide complementary improvements in the performance accuracy. Although averaging over more frames can yield additional performance gains, it requires collecting many frames within the channel’s geometric coherence time, which is impractical in high-mobility scenarios.\\
Moreover, increasing the number of frames leads to higher receiver latency. Therefore, averaging over five frames provides a practical trade-off between noise suppression and processing latency. Fig. \ref{frac effect} evaluates the performance of the proposed method under integer Doppler, fractional Doppler, and FDFD channel conditions. While fractional Doppler and FDFD induce DD domain leakage and sparsity loss, the proposed framework maintains stable performance by capturing the underlying DD domain structure. Further, to assess generalisation, a fractional Doppler mismatch between training and testing is introduced under the FDFD conditions, resulting in only a slight performance degradation. Fig. \ref{ber} shows that the BER performance with the SNR (PSNR $25$ dB) remains robust, despite fractional Doppler and FDFD conditions. This suggests that the proposed learning framework, aided by the guard structure, helps reduce the impact of fractional spreading.\\
Next, the computational complexity of the compared channel estimation schemes is discussed. For classical estimators, embedded pilot-based channel estimation involves thresholding over the effective DD grid of size $M_\tau \times N_\nu$, resulting in a complexity of $\mathcal{O}(M_\tau N_\nu)$ \cite{raviteja2018practical}. Whereas, raw DD domain observation and frame collection over $F$ pilot frames has complexity $\mathcal{O}(F M_\tau N_\nu)$. The OMP based method incurs a complexity of $\mathcal{O}(P M_\tau N_\nu)$, where $P$ defines the sparsity level of the DD domain channel. In \cite{he2023denoising}, the OMP estimate is further refined using a denoising convolutional neural network (DnCNN) denoiser. For DL-based estimators, computational complexity is deterministic and architecture-dependent; therefore, the number of parameters, FLOPs, and inference time are given in Table \ref{tab:csgmm_cc}.

\section{Conclusions}
\label{Conclusion} 
This treatise presents a DL based framework for channel estimation in OTFS systems, wherein the problem is reformulated as an image denoising problem. To ensure spectral efficiency and handle FDFD effects, a dedicated pilot structure is embedded within the data frame. Further, these DD domain observations are processed through a lightweight CNN-based image restoration framework based on NAFNet. The model not only performs denoising but also learns direct end-to-end learning of DD domain mappings. This enables efficient recover of on-grid channel structure and learning of off-grid energy dispersion, even under low-SNR conditions. Performance is further enhanced through frame-wise averaging of multiple denoised outputs, facilitating accurate CSI recovery under practical channel conditions.

\begin{table}[!t]
\scriptsize
\renewcommand{\arraystretch}{1.1}
\centering
\caption{ Model complexity and performance of DL based methods}
\label{tab:csgmm_cc}
\begin{tabular}{|l|c|c|c|}
\hline
\textbf{Method} & \textbf{No. of parameters} & \textbf{No. of FLOPs}& \textbf{Inference time}\\ \hline

\makecell[l]{\footnotesize 
Proposed} 
& $1.586 \times 10^{4}$
& $1.978 \times 10^{6}$ & $29.2734$ ms \\ \hline

\makecell[l]{\footnotesize SRCNN-based \\
\footnotesize FracNet \cite{guo2024otfs}} 
& $9.513 \times 10^{5}$
& $5.126 \times 10^{7}$ & $1083.73$ ms \\ \hline

\makecell[l]{\footnotesize FastDVDNet\\
\footnotesize (5-frames) \cite{jing2024learned}} 
& $1.083 \times 10^{5}$
& $5.308 \times 10^{6}$ & $54.6415$ ms \\ \hline

\makecell[l]{OMP$+$DnCNN \cite{he2023denoising}}
& $6.717 \times 10^{5}$ 
& $2.136{\times}10^{8}$ & $64.778$ ms \\ \hline
\end{tabular}
\end{table}

\bibliographystyle{IEEEtran}
\bibliography{bibliography.bib}
\end{document}